\documentclass[conference]{IEEEtran}
\IEEEoverridecommandlockouts
\usepackage{cite}
\usepackage{amsmath,amssymb,amsfonts}
\usepackage{algorithmic}
\usepackage{graphicx}
\usepackage{textcomp}
\usepackage{xcolor}
\usepackage{fixltx2e}
\def\BibTeX{{\rm B\kern-.05em{\sc i\kern-.025em b}\kern-.08em
    T\kern-.1667em\lower.7ex\hbox{E}\kern-.125emX}}
\begin{document}

\title{Environmental and Social Sustainability of Creative-Ai \\ {\large GenAICHI - Generative AI and CHI - 2022}}

\author{\IEEEauthorblockN{André Holzapfel}
\IEEEauthorblockA{\textit{KTH Royal Institute of Technology} \\
Stockholm, Sweden \\
holzap@kth.se}
\and
\IEEEauthorblockN{Petra Jääskeläinen}
\IEEEauthorblockA{\textit{KTH Royal Institute of Technology} \\
Stockholm, Sweden \\
ppja@kth.se}
\and
\IEEEauthorblockN{Anna-Kaisa Kaila}
\IEEEauthorblockA{\textit{KTH Royal Institute of Technology} \\
Stockholm, Sweden \\
akkaila@kth.se}
}

\maketitle

\begin{abstract}
The recent developments of artificial intelligence increase its capability for a creation of arts in both largely autonomous and collaborative contexts. In both contexts, Ai aims to imitate, combine, and extend existing artistic styles, and can transform creative practices. In our ongoing research, we investigate such Creative-Ai from sustainability and ethical perspectives. The two main focus areas are understanding the environmental sustainability aspects (material, practices) in the context of artistic processes that involve Creative-Ai, and ethical issues related to who gets to be involved in the creation process (power, authorship, ownership). This paper provides an outline of our ongoing research in these two directions. We will present our interdisciplinary approach, which combines interviews, workshops, online ethnography, and energy measurements, to address our research questions: How is Creative-Ai currently used by artist communities, and which future applications do artists imagine? When Ai is applied to creating art, how might it impact the economy and environment? And, how can answers to these questions guide requirements for intellectual property regimes for Creative-Ai?

\end{abstract}

\section{Introduction}

Recent development in artificial intelligence extends the possibilities for digital art industries beyond reproduction and distribution: Ai\footnote{The “i” is lowercase in Ai to emphasize the fact that the intelligence of current systems is quite different from human intelligence and has not yet reached a level of AGI (artificial general intelligence)} facilitates the commodification of automated creation of arts that aims at imitating, combining, and extending existing artistic styles (Creative-Ai), with a potential to transform the creative practices. This has major implications for artistic practice and society in general. We focus on studying these implications through questions, such as: how is creative-Ai currently used by creative communities, and what can guide the ethical practices for the design and use Creative-Ai? We study these questions through mixed-methods research that combines ethnography, digital methods, sustainability assessment, and recent human-computer interaction approaches (such as design fiction, diary studies, RtD-approaches to develop design tool-kits and heuristics for value-sensitive design) with a critical method. These studies will, then, motivate new approaches for the development of alternative directions for environmentally sustainable Creative-Ai, in which our research outcomes are expected to form ethical guidelines, theoretical knowledge, and material prototypes. We see these research directions fitting with the discussions of this workshop, as our research directly contributes to the ethics and sustainability discussions within the Generative Ai research community. 

In this workshop application, we outline our ongoing research in the context of our recently initiated project, and raise questions that we are currently exploring and would like to bring into the workshop discussions. The two main focus areas of our current works-in-progress are 1) environmental sustainability (material, practices) in the context of Ai arts and 2) ethical issues related to who is involved in the creation process (power, authorship, ownership). These two areas represent the core subjects of two individual PhD projects, respectively, and they are represented by the two PhD students co-authoring this paper.

 
\section{Environmental sustainability perspectives on Creative-Ai: material, practices}

Our main research directions related to environmental sustainability are to analyze the environmental impact of various Creative-Ai technologies and practices, and to use the outcome of the analysis to inform the design and use of Creative-Ai technologies. The environmental impact of Ai technologies without the contextualization of creative processes has been discussed by previous research \cite{schwartz_green_2020, strubell_energy_2019, lacoste_quantifying_2019, dhar_carbon_2020, ligozat_unraveling_2021}, with energy use and $CO^2$
emissions being analyzed by \cite{strubell_energy_2019, lacoste_quantifying_2019, dhar_carbon_2020, ligozat_unraveling_2021}. Strubell et al. \cite{strubell_energy_2019} have proposed energy and policy considerations for deep learning in general and Natural Language Processing (NLP) in particular. Others \cite{lacoste_quantifying_2019, dhar_carbon_2020} have instead focused on the impact of Ai on carbon emissions, for instance by estimating the carbon impact of different Graphics Processing Units (GPUs)\cite{lacoste_quantifying_2019}. However, none of the prior sustainability research specifically focuses on the use of Ai tools by the creative industries, nor on the context of art produced using Ai as a tool. Referring to the latter as \textit{Ai art}, we have argued in a recently submitted paper that the environmental footprint of Ai art needs to be studied to understand the environmental impact of the related artistic practices. Creative work processes are likely to differ from other Ai application areas, and the context of use should be understood so that we can develop approaches for more sustainable and ethical design and use of these Creative-Ai technologies. 

\subsection{Environmental impact of the current Creative-Ai practices} 
We conducted an online study aiming to understand how various creative practitioners use Creative-Ai in their practices. Our study revealed that some of the artists use massive data-sets for training the Ai. For instance, \cite{anadol} was using 45 terabytes of data – 587,763 image files, 1,880 video files, 1,483 metadata files, and 17,773 audio files equivalent of 40,000 hours of audio. Other artists collaborate with embodied Ai \cite{sougwen} in creation of the artworks. Regarding such human-Ai interaction within the Ai artistic practice, we should keep in mind that Ai materials are not necessarily static materials. Therefore, we have to not only account for the practice and agency of the artist, but potentially have to consider the agency of the Ai technology. This emerging interaction between these two agents necessitates a sustainability assessment that takes into account both material and process.

\begin{figure}[h]
\centering
\includegraphics[width=0.45\textwidth]{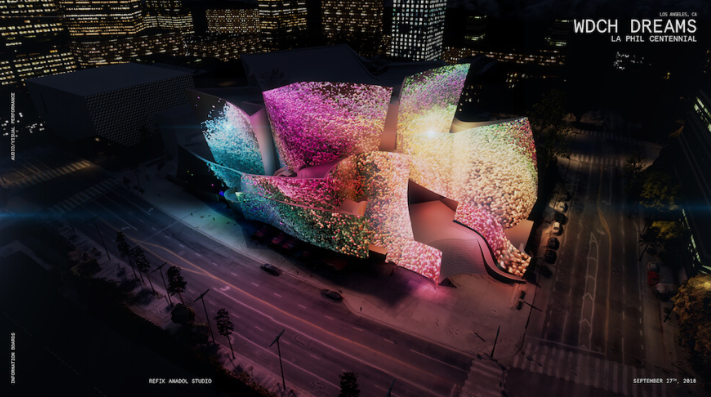}
\caption{Ai artwork of one of the artists we studied}
\end{figure}

Our insights gained so far imply that Ai materials are often used in the ideation, actualization and display phases of a creative process, following the process model proposed by \cite{botella_what_2018}. In the ideation phase, several artists seem to use Ai generators for inspiration and for helping ideas to emerge. In later stages like actualization, Ai is used by many artists to create the artwork itself. The Ai technologies can also be entangled with the display of the artwork: some artists are generating interactive artworks in real-time together with the audience or environments. 

\subsection{Understanding the environmental impact of Creative-Ai technologies}
We have measured energy usage in different kinds of running environments for VQGAN+Clip \cite{esser2020taming, radford2021learning}, as a popular generative art tool applied by many practitioners, and provided initial figures for these measurements in our paper that is currently under peer review. We are also currently expanding our mapping of the energy consumption on different kinds of devices and other algorithms/tools used by the creative communities (for example Google Magenta). These insights can help us to understand what ways of using the Creative-Ai technologies have high or low environmental impacts, and informs us how to design these technologies in the future. It will also inform the design of guidelines and toolkits for Creative-Ai practitioners.

\subsection{Design implications on system architecture}

One example of sustainability informed design is the design of latent spaces that serve the needs of the creative practitioners, but have minimal energy impact in terms of computation. We initiated a collaboration with another research group to explore the latent spaces of GAN algorithms for sound synthesis \cite{koray}. We expect that the results of this study will help us not to only understand the energy consumption and environmental impact of the AI algorithm development process, but will also shed light on the fidelity and accuracy required from these technologies in artistic practices. To the best of our knowledge, there is no prior research that combines sustainability assessment with the development of Ai technology for artistic purposes, and that conducts this development in close interaction with artists. We hope to be able to arrive a better understanding of the needs and demands by artists by conducting the described project. 

\subsection{Design implications on user interaction}

We also plan to explore the design of various Creative-Ai systems from a UX perspective, aiming to understand what kind of information is represented in the user interface to the users, and what kinds of interactions take place between the user and the system. Potential directions for future explorations are how to facilitate more sustainable ways of working with Ai in creative communities. Our question in this context is if a system can guide the users in terms of energy consumption, provide information of the energy consumption of their creative process, and potentially influence the creative process to reduce energy consumption. In either case, these design studies need to be rooted in the needs of the creative practitioners and artists, which is why in the beginning phases of the research we focus particularly on understanding these current practices and engaging in collaboration and discussions with creative practitioners to develop understanding of their needs. 

\subsection{Emerging Ai art ecosystem}
Sustainability issues are also related to the commercialization of Ai art. These include, for example, the use of emerging technologies such as blockchain, cryptocurrencies, and Non-fungible tokens (NFTs). An increasing number of Ai artists are exploring innovative high-tech pathways for selling their artworks - which in their turn will have environmental impacts. New initiatives emerge such as a "fair trade" art certificate that is offered to artists who use NFT's and cryptocurrencies for their works. These certificates indicate that a work of art has been produced in a socially sustainable manner, but currently there are no environmental sustainability certificates for Ai art. These initiatives and technologies are constituent factors as Ai art markets emerge. Since the value of art is co-created in society, the participating members determine what is considered valuable and appreciated, and in turn is in the focus of consumption. Ai art(s) shape new markets, which have the potential to introduce an additional environmental footprint through energy-heavy blockchain, cryptocurrencies, and NFT technologies. We have pointed this out in our latest study, and we intend to conduct further studies on environmental sustainability of Ai art markets based on such technologies.

\section{Social sustainability considerations of Creative-Ai: power, authorship, ownership}

Issues of power, control and ownership are entangled with Creative-Ai in several ways. In the midst of increasingly large networks of actors involved in the process of designing, implementing, training and maintaining the systems, the entire concept of authorship is becoming rather a collective than individual phenomenon. As some Ai systems gradually become more independent, the distinction between human and non-human as a creative agent starts to blur, which makes the situation more complicated both legally and ethically.

\subsection{Conjuction of law and ethics with creative Ai}

Collective and fluid definitions of authorship are in stark contrast with the legal framework of exclusive intellectual property rights. The outputs of the Creative-Ai system may under most copyright regimes  be protected if and only if the human fingerprint of individual human author(s) in the resulting work can be identified, and the threshold criteria of originality is satisfied \cite{ginsburg}. For many, if not the majority of Ai-created works, it is an open question whether such criteria of legal protection can be fulfilled \cite{drott}. 

Equally critical aspect is the data used as training material to shape an Ai model. Data mining legislation with regard to copyrighted works is relatively relaxed in the US, in certain Asian countries, and to a somewhat lesser extent in the EU. Yet, irrespective of the question whether or not the data used in the training was copyright-protected or in the public domain and therefore technically free for anyone to use, the data use itself can have serious neocolonialist implications for the creative communities that have contributed to this cultural corpus \cite{morreale}. Furthermore, such communities may be in more precarious positions than the system developers, with little or no voice to raise against unwarranted and potentially culturally insensitive uses of their works \cite{huang}. 

This illustrates the need for a sincere conversation of ethical issues related to the use of data in Creative-Ai. For this end, we are currently preparing a series of interviews and interactive workshops with Nordic Ai artists, exploring their encounters with these complex issues and their visions of what forms the future of ethically sustainable creative Ai could take. These observations of artistic processes will also provide empirical ground to inform the shaping of ethical guidelines for the field.

\subsection{Shortcomings of current ethics guidelines for Ai}
In the past decade, there has been a surge in the development of ethical frameworks that seek to identify central ethical themes relevant for general Ai use and development \cite{jobin}. As of the latest update in April 2020, AI Ethics Guidelines Global Inventory alone lists 173 guidelines or principles  developed by various governmental entities, NGOs or private companies \cite{inventory}. One prominent example is the Guidelines for Trustworthy AI by European Commission’s independent High-Level Expert Group on AI\cite{hleg}.  

As noted by several authors (\cite{ayling}, \cite{morley}) and the AI HLEG themselves\cite{altai}, the problem with these original frameworks was that they approach ethics from such a high conceptual level that they are unpractical for the creative professionals. As a solution to this problem, there is a growing need of applied ethics tools that would be more readily approachable in the development pipeline and more accessible for users with limited prior knowledge of ethics. 
There have been some attempts to address this need by translating the abstract ethics framework principles into a series of questions or discussion prompts. Some of the established approaches include the Open Data Institute's Data Ethics Canvas \cite{canvas} and the Ethical Toolkit for Engineering/Design Practice \cite{toolkit}, both of which take the approach of posing the development team an extended set of questions to elicit or expose issues of ethical relevance. More recent suggestions, such as the Data Ethics Decision Aid (DEDA) \cite{deda}, the Ethical Matrix \cite{matrix} and the ECCOLA Ai ethics card game \cite{eccola} distance themselves from what could be criticised as check-box ethics and aim instead towards more interactive analysis processes that encourage dialogue among participants. This could take the form of an elaborate workshop of exploring the ethics themes relevant for the project at hand, possibly facilitated by the tool providers as a paid service.   

\subsection{Prototypes for ethics-informed stakeholder analysis}

To address these issues from a pragmatic perspective, we are currently developing applied approaches that aim to help designers and developers explore the ethical and societal implications of their work with the focus specifically on the people directly and indirectly impacted. This is in line with the patiency-led ethics approach  \cite{coeck}. The focus of such work will be on tools or processes that are easily accessible both in terms of (free) platforms and structurally so that the team may also carry out the analysis without an external facilitator.  

The first prototype of such an approach was tested at the International Society for Music Information Retrieval Conference in 2021. Our ethics-informed stakeholder analysis starts with a collective brainstorming of various stakeholders that could be relevant for the scenario at hand on the Mentimeter platform\footnote{https://www.mentimeter.com/}. The participants then individually evaluate the identified stakeholders against one another on the axes of power and authorship, respectively. Combining these two planes of information, the stakeholders are mapped on a 2x2 matrix such that disparities of the power distribution between the stakeholder groups becomes more apparent. Finally, this matrix is used as a basis for discussing concrete action points that can be taken to alleviate the effects of power imbalances and make the design processes more diversely inclusive.  

This process will be further developed and tested in workshops planned for the early spring 2022. Some of the aspects under scrutiny include the axes of the analysis matrix and the question of how the specific ethics perspectives may be best integrated into the process. For example, the concept of 'authorship', which seemed to cause some confusion among the initial participants, could be replaced with the axis of 'interest', as used in the more traditional stakeholder analysis, which will change the nature of the task to a certain degree. Furthermore, in the next prototypes, the ethics perspectives will be more closely integrated into the analysis through the use of a limited number of focused question prompts that aim to widen the participants' brainstorming on who might be impacted by the development and use of creative Ai systems. 

\section{Summary}

Here, we have covered the directions of our on-going research within the domains of environmentally and socially sustainable Creative-Ai. For environmental sustainability these have included case studies on the process of Creative-Ai practices, quantifying the energy impact of the technologies used by creative practitioners, potential user-centered design implications for the system architecture and the user experience, and studies of the emerging Creative-Ai and Ai Arts ecosystems. For social sustainability, our research directions focus on the need to better understand the particularities of the artistic creation processes with Ai, as well as the practically motivated issue of stakeholder analysis. This way, we hope to close a gap between existing ethical frameworks and their actual implementation within creative communities and Creative-Ai developers. 

These research directions are of immediate relevance in the context of generative Ai, and with this workshop application we wish to participate in discussions and knowledge-exchange with researchers who work with generative Ai in various application contexts. We believe, that our research can enrich the research community's discussions on ethics and sustainability, and that we can learn from the specific challenges that researchers and practitioners in Generative-Ai community are encountering. 

\bibliographystyle{IEEEtran}
\bibliography{sample-base}

\end{document}